\documentclass[12pt,a4paper]{article}
\usepackage{xcolor}
\usepackage{jheppub}
\usepackage{epstopdf}
\usepackage{graphicx}
\usepackage{epsfig}
\usepackage{dcolumn}  
\usepackage{bm}    
\usepackage{amssymb} 
\usepackage{amsmath,bm}
\usepackage{amsfonts}    
\usepackage{slashed}  
\usepackage{youngtab}
\usepackage[mathscr]{euscript}
\usepackage{epsfig}
\usepackage{verbatim}

\title{
Gravity, Duality and Conformal Symmetry \footnote{Contribution to the Royal Society volume in celebration of Michael Duff's 70th birthday
}}

\author{Chris Hull} 
\affiliation{ The Blackett Laboratory, Imperial College London \\
\hspace*{0.3cm} Prince Consort Road London SW7 2AZ, U.K.}

\emailAdd{ c.hull@imperial.ac.uk}

\abstract{
The (4,0) supermultiplet in 6 dimensions contains a 4th rank tensor gauge field with the symmetries of the Riemann tensor and is superconformal, with 32+32 supersymmetries. Dimensional reduction on a circle gives the 5D N=8 supergravity multiplet, with the 4th rank tensor reducing to the graviton.
If there is an interacting (4,0) theory it should reduce to the full N=8 supergravity theory and so would give a conformal theory of gravity that would reduce to 
conventional gravity with  the usual 2-derivative action at low energies. This paper revisits the conjecture that  a non-lagrangian interacting (4,0) superconformal theory arises from a strong coupling limit  of 5D supergravity (suitably embedded in M-theory) describing    M-theory at  energies beyond the Planck scale. A key test for this is identified: M-theory toroidally compactified to 5D should have certain BPS states carrying a singlet central charge. These 1/2 BPS states are not related to any of the standard BPS states by dualities and do not correspond to non-singular soliton solutions -- they appear to correspond to singular solutions related to gravitational instantons. Such states are needed to provide the Kaluza-Klein modes for the compactified 6D theory. If there are no such states, then the conjecture is false, while  the presence of   such states would be strong indication that 
the conjecture could be true.}

\preprint{Imperial-TP-2022-CH-01}

\begin{document}

\maketitle

\section{ Symmetry Beyond the Planck Scale}

A lot of progress has been made in understanding quantum gravity at low
energies and small curvatures, but what happens at the Planck scale and beyond
remains something of a mystery. It is possible that a radically different
theory could emerge at the Planck scale and there have been many speculations
as to what might happen there.

One possibility is that, as one tries to probe distance scales smaller than
the Planck length, one could find that spacetime is no longer a continuum at
such scales and there is some discrete structure instead -- perhaps a matrix
model or some form of lattice theory.

Another speculation is that there could be a new highly symmetric phase
emerging at the Planck scale. An attractive proposal is that the fundamental
theory should be conformally invariant with no dimensionful parameters, with
the Planck scale arising via symmetry breaking -- see \cite{tHooft:2015vaz}  for a recent
discussion of this idea. A major obstacle to this is that the conformal
theories of a graviton field that we know of are all higher derivative, so
that although they have good ultra-violet properties, they have ghosts and it
is difficult to recover Einstein's theory from them. In string theory, it has
been proposed that at trans-Planckian energies there is a highly symmetric
phase with an infinite number of symmetries associated with an infinite number
of gauge fields becoming massless in some tensionless string limit. Evidence
for such a picture was found in high-energy scattering amplitudes by Gross and
Mende \cite{Gross:1987kza,Gross:1987ar}. More recently, it has been shown that for string theory in anti-de
Sitter space, tensionless strings arise when the curvature approaches a
critical value at the string scale giving an infinite set of symmetries and
signalling a transition to a new phase \cite{Gaberdiel:2017oqg,Giribet:2018ada,Gaberdiel:2018rqv,Eberhardt:2018ouy,Gaberdiel:2021qbb}.

The standard picture of quantum gravity is that it arises from the
quantsation of a massless spin two field, and this field is then interpreted
as the metric governing the Riemannian geometry of spacetime. However, it is
interesting to ask whether a massless spin two field provides the only way of
formulating gravity. At low energies, we would require that the theory can be
viewed as a theory of a graviton. However, this theory could have an
ultra-violet origin that is a theory of another field, perhaps in higher
dimensions.

This paper will revisit the conjecture of \cite{Hull:2000zn} in which it was proposed that
gravity could be conformally invariant but formulated not in terms of the
familiar spin-two graviton field but in terms of something more exotic. In the
free limit, the theory of \cite{Hull:2000zn} is formulated in terms of a four index gauge
field $C_{M N P Q}$ (with the symmetries of the Riemann tensor), but on
compactification on a circle, a conventional ghost-free theory of a graviton
emerges at scales small compared to the compactification scale, which also
sets the Planck scale in the compactified theory. The outstanding question is
whether there could be an interacting version of this theory that reduces to
general relativity or supergravity.

This conjecture arose from supersymmetry. There is a remarkable supermultiplet
in 6 dimensions with (4,0) supersymmetry which is fact a superconformal
multiplet so that the theory has 32 normal supersymmetries plus 32 conformal
supersymmetries. It has the gauge field $C_{M N P Q}$ together with 27
self-dual 2-forms and 42 scalars. Reducing on a circle gives 5-dimensional N=8
supergravity. It exists as a free theory, but again it is not known whether
there is an interacting version. If there is, it would be expected to be a
non-lagrangian theory of the kind arising in the (2,0) theory that arises as
the world-volume theory of the M5-brane. An interacting theory would be very
interesting, giving a conformal theory of gravity that reduces to a
conventional theory of gravity at low energies. Little progress has been made
on this proposal since it was made in 2000, partly because it was waiting for
progress in understanding the (2,0) theory. Now much more is known about the
(2,0) theory, it is perhaps time to revisit this conjecture. In this paper the (4,0) theory will be reviewed following \cite{Hull:2000zn,Hull:2000rr,Hull:2000ih,Hull:2001iu}  
and the  conjectures concerning it \cite{Hull:2000zn} will be re-examined.

\section{The 6D (2,0) theory \& 5D Yang-Mills}

There is considerable evidence that there exists an interacting (2,0)
non-lagrangian superconformal field theory (SCFT) in six spacetime dimensions \cite{Berkooz:1997cq,Seiberg:1997ax,Aharony:1997th}.
Moreover, this provides a key to understanding strongly-coupled maximally
supersymmetric Yang-Mills (SYM) theory in D=5 \cite{Witten:1995zh,Rozali:1997cb,Lambert:2010iw,Douglas:2010iu} and S-duality in
D=4 \cite{Witten:1995zh},\cite{Witten:2009at}.

The free (2,0) supersymmetric theory in 6-D has a 2-form gauge field $B$
whose field strength $H = dB$ is self-dual, $H = \ast H $. The other fields are
5 scalars transforming as a $\bf{5}$ under the R-symmetry group
${Sp} (2)$ and 4 symplectic Majorana-Weyl fermions transforming as a
$\bf{4}$ under ${Sp} (2)$.

It is known that there is a quantum interacting (2,0) SCFT, but it is not
clear how to formulate the interacting theory. It is clearly not a
conventional theory of gauge fields -- it is said to be non-lagrangian -- but
nonetheless it is known that its dimensional reduction to 5 or less dimensions
does give a Yang-Mills gauge theory. Dimensional reduction of the free (2,0)
theory on a circle gives 5-D N=4 abelian gauge theory with $B$ giving rise to
the 5-D vector potential $A$. The interacting (2,0) SCFT is non-lagrangian but
it reduces to a conventional field theory, 5D N=4 SYM.

The (2,0) theory arises as the strong coupling limit of 5D SYM \cite{Witten:1995zh,Rozali:1997cb,Lambert:2010iw,Douglas:2010iu}. The
interacting (2,0) theory compactified on a circle of radius $R$ gives 5D SYM
with coupling constant $g_{YM}$ with $g_{YM}^2 = R$, so that as $g_{YM}
\rightarrow \infty$ the radius $R \rightarrow \infty$ and at strong coupling
the circle decompactifies to give a 6D theory. The 5D SYM theory is
non-renormalizable and is known to have divergences \cite{Bern:2012di} so here it is regarded
as being embedded in some UV complete theory, such as string theory. The (2,0)
theory also arises as a decoupling limit of the world-volume theory of a stack
of M5-branes and in IIB string theory compactified on K3.

The 5D SYM theory has BPS 0-branes given by lifting self-dual Yang-Mills
instantons in 4 Euclidean dimensions to soliton world-lines in 4+1 dimensions,
taking the product of the 4D instanton solution with a timelike line. These
solitons have mass $M \propto |n| / g_{YM}^2$ where $n$ is the instanton
number, so that these become light as the dimensionful Yang-Mills coupling
$g_{YM}$ becomes large. These solitons should be interpreted as Kaluza-Klein
modes for a $D = 6$ theory compactified on a circle of radius $R = g_{YM}^2$
 so that in the strong coupling limit an extra dimension opens up to give a
6-dimensional (2,0) supersymmetric theory \cite{Witten:1995zh}. In the abelian theory, the 5D
vector gauge field is replaced by a 6D 2-form gauge field $B_{MN}$ with
self-dual field strength, and the 5 scalar fields are all promoted to scalar
fields in 6 dimensions

The $D = 6$ theory is believed to be a non-trivial superconformally invariant
quantum theory \cite{Seiberg:1997ax} and the $D = 5$ gauge coupling arises as the radius of the
compactification circle, $g_{YM}^2 = R$ \cite{Witten:1995zh}. The relationship between the $D = 5$
and $D = 6$ theories is straightforward to establish for the free case in
which the Yang-Mills gauge group is abelian, but in the interacting theory the
6-dimensional origin of the $D = 5$ non-abelian interactions is mysterious;
there are certainly no local covariant interactions that can be written down
that give Yang-Mills interactions when dimensionally reduced \cite{Bekaert:2000qx}. Nonetheless,
the fact that these $D = 5$ and $D = 6$ theories arise as the world volume
theories of D4 and M5 branes respectively  gives strong support for the
existence of such a 6-dimensional origin for the gauge interactions.

The W-bosons and magnetic strings in $D = 5$ arise from self-dual strings in
$D = 6$. At the origin of moduli space the W-bosons become massless and the
tensions of the self-dual strings in $D = 6$ must also become zero. The nature
of the theory at such points is unclear. Nonetheless, given that mysterious
interactions with no conventional field theory formulation arise in the
M5-brane world-volume theory, it is natural to seek similar unconventional
interactions elsewhere in M-theory.

\section{The 6D (4,0) theory \& 5D Supergravity}

Given the close relationship between gauge theory and gravity, it is natural
to ask whether there could be a story for gravity similar to that of the
relation between the 6D (2,0) theory and 5D SYM. Remarkably, there is a free
SCFT with (4,0) supersymmetry in 6-D, and if this has an extension to an
interacting theory, it would provide an exotic conformal theory giving
supergravity in D<6 and could arise as a strong coupling limit of 5D
supergravity.

Five-dimensional $N = 8$ supergravity (ungauged) has a global $E_6$ symmetry
and a local ${Sp} (4) = {USp} (8)$ R-symmetry. It is
non-renormalisable, and will be regarded as arising as a massless sector of
some consistent theory, such as M-theory compactified on a 6-torus, in which
the global $E_6$ symmetry is broken to a discrete U-duality subgroup \cite{Hull:1994ys}.
The massless bosonic fields consist of a graviton, 27 abelian vector fields
and 42 scalars. The dimensional coupling constant is the 5-dimensional Planck
length $l$, so that strong gravitational coupling is the limit as $l \to
\infty$. For such a limit it is natural to look first for a theory with 32
supersymmetries and ${Sp} (4)$ R-symmetry and to expect that BPS states
are protected and survive as the coupling $l$ is increased.

The multiplet has (4,0) supersymmetry in $D = 6$ and was studied in \cite{Hull:2000zn}.
The particle content was identified in \cite{Strathdee:1986jr} and in \cite{Hull:2000zn} the covariant field content and gauge symmetry was found and further studied in \cite{Hull:2000rr,Hull:2000ih,Hull:2001iu}. Actions  for the free theory have been  considered in \cite{Henneaux:2017xsb,Bertrand:2020nob,Bertrand:2022pyi} and further discussion of the theory can be found in \cite{Schwarz:2000zg,Minasian:2020vxn,Gunaydin:2020mod,Lekeu:2021oti,Cederwall:2020dui}.

Instead of a graviton, it has an exotic fourth-rank tensor gauge field $C_{MN
PQ}$ with the algebraic properties of the Riemann tensor
\begin{equation}
  C_{M N P Q} = - C_{N M P Q} = - C_{M N Q P} = C_{P Q M N}
\end{equation}
and the gauge symmetry
\begin{equation}
  \delta C_{MN \hspace{0.17em} PQ} = \partial_{[M} \chi_{N] PQ} +
  \partial_{[P} \chi_{Q] MN} - 2 \partial_{[M} \chi_{NPQ]}  \label{delcis}
\end{equation}
with parameter $\chi_{MPQ} = - \chi_{MQP}$. The invariant field strength is
\begin{equation}
  G_{MNP \hspace{0.17em} QRS} = \frac{1}{36}  (\partial_M \partial_S C_{NP
  \hspace{0.17em} RS} + \ldots) = \partial_{[M} C_{NP] \hspace{0.17em} [QR,
  S]}
\end{equation}
and in the (4,0) multiplet it satisfies the self-duality constraint
\begin{equation}
  G_{MNP \hspace{0.17em} QRS} = \frac{1}{6} \epsilon_{MNPTUV} G^{TUV}  _{QRS}
\end{equation}
or $G = \ast G$. In addition, there are 27 2-form gauge-fields with self-dual 3-form
field strengths, 42 scalars, 48 symplectic Majorana-Weyl fermions and, instead
of gravitini, 8 spinor-valued 2-forms $\Psi_{MN}^a$ which satisfy a symplectic
Majorana-Weyl constraint and have self-dual field strengths. The fermionic
gauge symmetry is of the form
\begin{equation}
  \delta \Psi_{MN}^a = \partial_{[M} \varepsilon_{N]}^a
\end{equation}
with parameter a spinor-vector $\varepsilon_N^{\alpha a}$. The free theory
based on this multiplet is a superconformally invariant theory, with conformal
supergroup ${OSp}^{\ast} (8 / 8)$  \cite{Hull:2000zn,Chiodaroli:2011pp}. This has bosonic subgroup
${USp} (8) \times {SO}^{\ast} (8) = {Sp} (4) \times {SO}
(6, 2)$ and 64 fermionic generators, consisting of the 32 supersymmetries of
the (4,0) superalgebra and 32 conformal supersymmetries.

It is remarkable that, in going from $D = 5$ to $D = 6$ in the free theory,
the vector gauge fields $A_{\mu}$ are lifted to 2-forms $B_{MN}$, the
gravitini $\psi_{\mu}$ are lifted to spinor-valued 2-forms $\Psi_{MN}$ and the
graviton $h_{\mu \nu}$ is lifted to the gauge field $C_{MN P Q}$, with these
$D = 6$ gauge-fields all satisfying self-duality constraints. Electrically
charged 0-branes and magnetic strings in $D = 5$ lift to BPS self-dual strings
in $D = 6$. In \cite{Hull:2000zn} it was shown that the dimensional reduction of the
free (4,0) theory on a circle indeed gives the linearised $D = 5, N = 8$
supergravity theory, with gravitational coupling (Planck length) given by the
circle radius $l = R$. However, there are no covariant local interactions in
$D = 6$ for this multiplet that could give rise to the $D = 5$ supergravity
interactions.

There is then a close analogy between the $D = 5, N = 4$ Yang-Mills theory and
the $D = 5, N = 8$ supergravity. The linearised versions of these theories
both arise from the dimensional reduction of a free superconformal field
theory in $D = 6$, with the dimensional $D = 5$ coupling constant arising from
the radius of compactification, so that the strong coupling limit of the $D =
5$ free theories is a decompactification to $D = 6$. For the interacting $D =
5$ Yang-Mills theory, there are a number of arguments to support the
conjecture that its strong coupling limit should be an interacting
superconformal theory in $D = 6$ with (2,0) supersymmetry, even though such a
theory has not been constructed directly and indeed cannot have a conventional
field theory formulation. This led to the conjecture of \cite{Hull:2000zn} that the
situation for $D = 5$ supergravity is similar to that of $D = 5$
super-Yang-Mills, and that a certain strong coupling limit of the interacting
supergravity theory should give an interacting theory whose free limit is the
(4,0) theory in $D = 6$. In this case there is no analogue of the M5-brane
argument to support this, although the M5-brane case does set a suggestive
precedent.

The limiting theory should have some novel form of interactions which give the
non-polynomial supergravity interactions on reduction. It could be that these
are some non-local or non-covariant self-interactions of the (4,0) multiplet,
or it could be that other degrees of freedom might be needed; one candidate
might be some form of string field theory. However, if a strong coupling limit
of the theory does exist that meets the requirements assumed here, then the
limit must be a (4,0) theory in six dimensions, and this would predict the
existence of interactions arising from the strong coupling limit of the
supergravity interactions. Although it has not been possible to prove the
existence of such a limit, it is remarkable that there is such a simple
candidate theory for the limit with so many properties in common with the
(2,0) limit of the $D = 5$ gauge theory. Conversely, if there is a
6-dimensional phase of M-theory which has (4,0) supersymmetry, then its circle
reduction to $D = 5$ must give an $N = 8$ supersymmetric theory and the
scenario described here should apply.

In the context in which the $D = 5$ supergravity is the massless sector of
M-theory on $T^6$, then the $D = 6$ superconformal theory would be a field
theory sector of a new 6-dimensional superconformal phase of M-theory. This
could tell us a great deal about M-theory: it would be perhaps the most
symmetric phase of M-theory so far found, with a huge amount of unbroken gauge
symmetry, and would be a phase that is not well-described by a conventional
field theory at low energies, so that it could give new insights into the
degrees of freedom of M-theory. 

\section{The Superalgebra and BPS States}

The five-dimensional $N = 2 n$ superalgebra with automorphism group ${Sp}
(n)$ is
\begin{eqnarray}
  \{Q_{\alpha}^a, Q_{\beta}^b \} & =& \Omega^{ab} 
  (\Gamma^{\mu} C)_{\alpha \beta} P_{\mu} + \Omega^{ab} C_{\alpha \beta} K
  \nonumber\\
  & +& (\Gamma^{\mu} C)_{\alpha \beta} Z_{\mu}^{ab} + C_{\alpha \beta} Z^{ab}
  + \frac{1}{2}  (\Gamma^{\mu \nu} C)_{\alpha \beta} Z_{\mu \nu}^{ab}   \label{salg}
\end{eqnarray}
where $\mu, \nu = 0, 1, \ldots, 4$ are spacetime indices, $\alpha = 1, \ldots,
4$ are spinor indices, $a = 1, \ldots, N$ are $Sp (n)$ indices, $C_{\alpha
\beta}$ is the charge conjugation matrix and $\Omega^{ab}$ is the symplectic
invariant of $Sp (n)$. The supercharges $Q_{\alpha}^a$ are symplectic Majorana
spinors satisfying
\[ (\bar{Q})^{\alpha}_a = C^{\alpha \beta} \Omega_{ab} Q_{\beta}^b \]
For the $D=5$, $N = 4$ super-Yang-Mills theory, the charges can be identified as follows \cite{Lambert:2010iw}.
The 5 central charges $Z^{ab} = -
Z^{ba}$ with $\Omega_{a b} Z^{ab} = 0$ are proportional to the five electric
charges $q^I \propto \int tr (\ast F \phi^I)$ and are carried by massive
vector multiplets. There is no vector field coupling to the singlet central
charge $K$, which is a topological charge carried by the instantonic 0-branes
and is proportional to the instanton number. There is a corresponding
conserved topological current
\begin{equation}
  j = \ast {tr} (F \wedge F) \label{inst}
\end{equation}
The spatial components $Z_i^{ab}$ ($i, j = 1, ..., 4$) of $Z_{\mu}^{ab}$ are
the magnetic charges carried by the magnetically charged strings. The other
charges on the right-hand-side of   (\ref{salg}) are 2,3 and 4-brane charges.  For
example, a vortex solution in 2+1 dimensions lifts to a 2-brane in 4+1
dimensions with charge $Z_{ij}$.

The $N = 8$ superalgebra has automorphism group ${Sp} (4)$ and is given
by   (\ref{salg})  with $a, b = 1, \ldots, 8$. The algebra with scalar central charges only
is
\begin{equation}
  \{Q_{\alpha}^a, Q_{\beta}^b \} = \hspace{0.17em} \Omega^{ab}  (\Gamma^{\mu}
  C)_{\alpha \beta} P_{\mu} + C_{\alpha \beta}  (Z^{ab} + \Omega^{ab} K)
\end{equation}
The 27 central charges $Z^{ab}$ satisfy $Z^{ab} = - Z^{ba}$, $\Omega_{a b}
Z^{ab} = 0$, and are the electric charges for the 27 vector fields (dressed
with scalars). There are 28 central charges but only 27 vector fields, and $K$
is not the conserved charge for any gauge field. However, on dimensional
reduction from 5 to 4 dimensions, $K$ becomes one of the 28 magnetic charges
of $D = 4, N = 8$ supergravity, and is the one coupling to the gravi-photon
field (i.e. the electromagnetic field from the dimensional reduction of the
metric) so that it is the Kaluza-Klein monopole charge. U-duality requires
that 1/2-supersymmetric states with $M = |K|$ occur in the $D = 4$ BPS
spectrum, and $K$ is quantized. A key question is whether there are
1/2-supersymmetric states in $D = 5$ carrying the central charge $K$ with $M =
|K|$. Presumably $K$ should be quantized, so that $K \propto n / l$ where $n$
is an integer. In any case, there is a conserved current analogous to  (\ref{inst})  given
by
\begin{equation}
  j = \ast {tr} (R \wedge R)
\end{equation}

The (4,0) theory has 27 self-dual 2-forms and these couple to 27 self-dual BPS
strings, or more precisely, to strings whose charges take values in a
27-dimensional lattice. In addition, there are 42 scalars and these can couple
to BPS 3-branes in 6 dimensions.

The (4,0) superalgebra in six dimensions with central charges is
\begin{eqnarray}
  \{Q_{\alpha}^a, Q_{\beta}^b \} & = \hspace{0.17em} \Omega^{ab}  (\Pi_+
  \Gamma^M C)_{\alpha \beta} P_M + (\Pi_+ \Gamma^M C)_{\alpha \beta} Z_M^{ab}
  \nonumber\\
  & + \frac{1}{6}  (\Pi_+ \Gamma^{MNP} C)_{\alpha \beta} Z_{MNP}^{ab} 
\end{eqnarray}
where $\Pi_+$ is the chiral projector
\[ \Pi_{\pm} = \frac{1}{2}  (1 \pm \Gamma^7) \]
The 27 one-form charges satisfy $Z_M^{ab} = - Z_M^{ba}$, $Z_M^{ab} \Omega ab =
0$, while $Z_{MNP}^{ab} = Z_{MNP}^{ba}$ and is a self-dual 3-form,
\begin{equation}
  Z_{MNP}^{ab} = \frac{1}{6} \epsilon_{MNPQRS} Z^{ab \hspace{0.17em} QRS} \label{sdu}
\end{equation}
The 27 charges $Z_i^{ab}$ with spatial indices $i, j = 1, \ldots, 5$ are the
string charges, and $Z_{ijk}^{ab}$ are the 36 3-brane charges, while
$Z_0^{ab}$ are the charges for space-filling 5-branes. (Note that the
self-duality condition  (\ref{sdu})  implies that $Z_{0 ij}$ are not independent charges.)

On dimensional reduction to 5 dimensions, the charges decompose as \cite{Hull:2000rr}
\begin{eqnarray}
  Z_M^{ab} & \to & (Z_{\mu}^{ab}, Z_5^{ab} = Z^{ab}) \nonumber\\
  P^M & \to & (P^{\mu}, P^5 = K) \nonumber\\
  Z_{MNP}^{ab} & \to & Z_{\mu \nu 5}^{ab} = Z_{\mu \nu}^{ab} 
\end{eqnarray}
The extra component of momentum becomes the $K$-charge, so that the
5-dimensional instantonic 0-brane is, from the $D = 6$ viewpoint, a wave
carrying momentum $P^5$ in the extra circular dimension. The electric 0-branes
in $D = 5$ arise from strings winding around the 6th dimension while the
magnetic strings arise from unwrapped $D = 6$ strings. The 2-branes and
3-branes in $D = 5$ arise from 3-branes in $D = 6$, while the space-filling $D
= 5$ 4-branes come from wrapped $D = 6$ 5-branes.

\section{Strong Coupling Limits and Solitons}

The 5D SYM theory has BPS solitons carrying the central charge $K$ in the
superalgebra  (\ref{salg}). These arise from self-dual or anti-self-dual YM instantons in
4 Euclidean dimensions lifted to 0-branes in 5D by taking a product of the 4D
instanton with time. They have mass
\begin{equation}
  M \propto \frac{|n|}{g_{YM}^2}
\end{equation}
$\text{}$where $n$ is the instanton number, so that these become light as the
dimensionful Yang-Mills coupling $g_{YM}$ becomes large. These are interpreted
as Kaluza-Klein modes for a $D = 6$ theory compactified on a circle of radius
$R = g_{YM}^2$, so that the strong coupling limit is a decompactification
limit $R \rightarrow \infty$. The current $j = \ast {tr} (F \wedge F)$
represents the density of these solitons. The BPS states carrying the charge
$K$ fit into massive 5D supermultiplets with precisely the right structure to
be the KK modes for a (2,0) theory \cite{Hull:2000cf}. In particular, they have massive
self-dual 2-form fields $B$ with mass $m$ satisfying
\begin{equation}
  {dB} = \pm m \ast B
\end{equation}
together with massive fermions and scalars.

The D=5 SYM is non-renormalizable but a UV completion can be defined within
string theory e.g. the D4-brane theory. The strong coupling limit is then
defined within string theory e.g. the strong coupling limit of multiple D4
branes gives multiple M5 branes.

As the coupling $g_{Y M}$ is dimensionful, the limit is better expressed as a
high energy limit, going to energies $E$ large compared to the YM scale,
\begin{equation}
  E \gg \frac{1}{g_{Y M}^2}
\end{equation}
so that at ultra-high energies the extra circle dimension becomes observable.

In the free case, the dimensional reduction can be done explicitly and the
free (2,0) theory reduced on a circle gives abelian SYM in 5D. However, in the
abelian case there are no non-singular instantons and so no solitonic
0-branes, but the Kaluza Klein modes can be associated with singular zero-size
instantons. A similar issue arises for the non-abelian SYM with the gauge
symmetry spontaneously broken to an abelian subgroup, with the instantonic
solitons shrinking to zero size as the Higgs expectation value is turned on.
However, in this case the soliton can be stabilized by turning on an electric
charge to give a regular BPS soliton solution \cite{Lambert:1999ua}, so that there should be BPS
0-branes in the 5D theory even when the symmetry becomes abelian.

The free (4,0) theory compactified on a circle of radius $R$ gives 5D
linearised supergravity with 5D Planck length $l = R$ plus Kaluza-Klein modes
with mass
\begin{equation}
  M \propto \frac{|n|}{R} = \frac{|n|}{l}  \label{spec}
\end{equation}
for integers $n$.

For the interacting 5D N=8 supergravity theory (embedded in string theory to
give a UV completion, as in M-theory compactified on a 6-torus), a necessary
condition for it to decompactify to a 6D (4,0) theory is that it should have
1/2-BPS states carrying the charge $K$ with a spectrum of the form  (\ref{spec}). From \cite{Hull:2000cf},
the 1/2-BPS states carrying $K$ fit into short 5D multiplets whose highest
spin field is a massive self-dual field $C_{\mu \nu \rho \sigma}$ satisfying
\begin{equation}
  \partial_{[\mu } C_{\nu  \rho] \sigma \tau} = \pm
  \frac{1}{2} m \epsilon_{\mu \nu \rho} ^{\alpha \beta} C_{\alpha \beta \sigma
  \tau}
\end{equation}
as required for the KK modes of $C_{M N {PQ}}$.

The key question is then whether there are BPS states carrying the charge
$K$, and if so, what is their spectrum. This then provides a crucial test for the conjecture  that there is a 6D (4,0) theory arising as a limit of M-theory.
M-theory toroidally compactified to 5D should have certain BPS states carrying the singlet central charge $K$. If there are no such states, then the conjecture is false, while  the presence of   such states would be strong indication that 
the conjecture could be true.

Independent of this conjecture, it is interesting in
any case to ask whether there are BPS states carrying $K$. All other central
charges in the 5D superalgebra are carried by BPS states and so it would be
strange if there were none carrying $K$. The charge $K$ is a singlet under the
$E_6 (\mathbb{Z})$ U-duality and the ${Sp} (4)$ R-symmetry so such BPS
states would not be related to any of the standard BPS branes by duality.

The explicit form for $K$ can be found from the superalgebra and was given in
\cite{Hull:1997kt}. If the 5D theory is compactified on a circle to 4D, then there are BPS
states carrying $K$ which are the KK monopoles, with $K$ related to the NUT
charge and hence the 4D magnetic charge \cite{Hull:1997kt}.

Just as for 5D SYM, for supergravity taking the product of a self-dual
gravitational instanton in 4 Euclidean dimensions with a timelike line gives a
1/2-supersymmetric configuration associated with the charge $K$. It would be
desirable to have such a solution which is asymptotic to 5D Minkowski space
but unfortunately there are no such non-trivial non-singular solutions. There
are no asymptotically Euclidean gravitational instantons but there are ALE
ones which are asymptotic to a discrete quotient of 4D Euclidean space (See e.g. \cite{Eguchi:1980jx}). Thus
there are no regular solitons with the desired asymptotics. However, this does
not preclude the existence of BPS states of this kind: they could be
associated with singular solutions or zero-size gravitational instantons. 
The role of gravitational instantons is further suggested by the topological symmetry associated with 
the current $j = \ast {tr} (R \wedge R)$ associated with the gravitational instanton density.
It
is interesting that there remains a possibility of further BPS states in
M-theory that are not dual to any of the known branes and which correspond to
singular supergravity solutions.

\section{Gravity and Geometry}

Particularly intriguing are the consequences for gravity. In $D = 5$, gravity
is described geometrically in terms of a metric $g_{\mu \nu}$, but at strong
coupling it is described instead in terms of the gauge field $C_{MNPQ}$ in $D
= 6$ (at least in the free case). This suggests the possibility of some new
structure which reduces to Riemannian geometry but which is more general and
is the appropriate language for describing gravity beyond the Planck scale.
For example, while $g_{\mu \nu}$ provides a norm for vectors and a notion of
length, $C_{MNPQ}$ could provide a norm $C_{MNPQ} \omega^{MN} \omega^{PQ}$ for
2-forms $\omega^{MN}$ and hence gives a notion of area that is not derived
from a concept of length \cite{Hull:2000ih}. It may be that the interacting theory is not
described in conventional $D = 6$ spacetime at all, but in some other arena.

There seem to be three main possibilities. The first is that there is no
interacting version of the (4,0) theory, that it only exists as a free theory,
and that the limit proposed in\cite{Hull:2000zn} only exists for the free $D = 5$
theory. The second is that an interacting form of the theory does exist in 6
spacetime dimensions, with $D = 6$ diffeomorphism symmetry. The absence of a
spacetime metric means that such a generally covariant theory would be of an
unusual kind. The third and perhaps the most interesting possibility is that
the theory that reduces to the interacting supergravity in $D = 5$ is not a
diffeomorphism-invariant theory in six spacetime dimensions, but is something
more exotic.

For gravity in any dimension, the full non-linear gauge symmetry is
\begin{equation}
  \delta g_{\mu \nu} = 2 \nabla_{(\mu} \xi_{\nu)}  \label{dif}
\end{equation}
If the metric is written as
\begin{equation}
  g_{\mu \nu} = \bar{g}_{\mu \nu} + h_{\mu \nu} 
\end{equation}
in terms of a fluctuation $h_{\mu \nu}$ about some background metric
$\bar{g}_{\mu\nu}$ (e.g. a flat background metric) then two main types of
symmetry emerge. The first consists of \lq background reparameterizations'
\begin{equation}
  \delta \bar{g}_{\mu \nu} = 2 \bar{\nabla}_{(\mu } \xi_{\nu)},
  \qquad \delta h_{\mu \nu} =\mathcal{L}_{\xi} h_{\mu \nu}   \label{bac}
\end{equation}
where $\bar{\nabla}$ is the background covariant derivative with connection
constructed from $\bar{g}_{\mu \nu}$, while $h_{\mu \nu}$ transforms as a
tensor ($\mathcal{L}_{\xi}$ is the Lie derivative with respect to the vector
field $\xi$), as do all other covariant fields. The second is the \lq gauge
symmetry' of the form
\begin{equation}
  \delta \bar{g}_{\mu \nu} = 0, \qquad \delta h_{\mu \nu} = 2 \nabla_{(\mu}
  \zeta_{\nu)}   \label{gag}
\end{equation}
in which $h_{\mu \nu}$ transforms as a gauge field and the background is
invariant. There is in addition the standard shift symmetry under which
\begin{equation}
  \delta \bar{g}_{\mu \nu} = \alpha_{\mu \nu}, \qquad \delta h_{\mu \nu} = -
  \alpha_{\mu \nu}
  \label{shift}
\end{equation}
In terms of the full metric $g_{\mu \nu}$, there is no shift symmetry and a
unique gauge symmetry  (\ref{dif}); the various types of symmetry (\ref{bac}),(\ref{gag}),(\ref{shift})   are an artifice of
the background split. The shift symmetry is a signal of background
independence and plays an important role in the interacting theory.

The linearised $D = 5$ supergravity theory has both background
reparameterization and gauge invariances given by the linearised forms of  (\ref{bac}),(\ref{gag})  
respectively, and both of these have origins in $D = 6$ symmetries of the free
(4,0) theory. The background reparameterization invariance lifts to the
linearised $D = 6$ background reparameterization invariance
\begin{equation}
  \delta \bar{g}_{MN} = 2 \partial_{(M} \xi_{N)}, \qquad \delta C_{MNPQ}
  =\mathcal{L}_{\xi} C_{MNPQ}
\end{equation}
with the transformations leaving the flat background metric $\bar{g}_{MN}$
invariant forming the $D = 6$ Poincar{\'e} group. The $D = 5$ gauge symmetry
given by the linearised form of  (\ref{gag})    arises from the $D = 6$ gauge symmetry  (\ref{delcis})  
with $\delta \bar{g}_{MN} = 0$ and the parameters related by $\zeta^{\mu} =
\chi^{55 \mu}$. The $D = 6$ theory has no analogue of the shift symmetry, and
the emergence of that symmetry on reduction to $D = 5$ and dualising to
formulate the theory in terms of a graviton $h_{\mu \nu}$ comes as a surprise
from this viewpoint.

The gravitational interactions of the full supergravity theory in $D = 5$ are
best expressed geometrically in terms of the total metric $g_{\mu \nu}$. If an
interacting form of the (4,0) theory exists that reduces to the $D = 5$
supergravity, it must be of an unusual kind. One possibility is that there is
no background metric of any kind in $D = 6$, and the full theory is formulated
in terms of a total field corresponding to $C$, with a spacetime metric
emerging only in a particular background $C$ field and a particular limit
corresponding to the free theory limit in $D = 5$.

It is not even clear that the interacting theory should be formulated in a $D
= 6$ spacetime. In $D = 5$, the diffeomorphisms act on the coordinates as
\begin{equation}
  \delta x^{\mu} = \xi^{\mu}
\end{equation}
In the (4,0) theory, the parameter $\xi^{\mu}$ lifts to a parameter
$\chi^{MNP}$. If the coordinate transformations were to lift, it could be to
something like a manifold with coordinates $X^{MNP}$ transforming through
reparameterisations
\begin{equation}
  \delta X^{MNP} = \chi^{MNP}
\end{equation}
with the $D = 5$ spacetime arising as a submanifold with $x^{\mu} = X^{55
\mu}$. Another possibility is as follows. The diffeomorphism $\delta x^{\mu} =
\xi^{\mu}$ gives a gauge transformation for the graviton

\begin{equation}
  \delta g_{\mu \nu} = \partial_{(\mu } \xi_{ \nu)} +
  \cdots
\end{equation}
with the index on the parameter lowered with the metric
\begin{equation}
  \xi_{\mu} = \xi^{\nu} g_{\mu \nu}
\end{equation}
The parameter of the gauge transformation
\begin{equation}
  \delta C_{MN \hspace{0.17em} PQ} = \partial_{[M} \chi_{N] PQ} +
  \partial_{[P} \chi_{Q] MN} - 2 \partial_{[M} \chi_{NPQ]}
\end{equation}
could be related to that of a 6D diffeomorphism
\begin{equation}
  \delta X^M = \xi^M
\end{equation}
by
\begin{equation}
  \chi_{M N P} = \xi^Q C_{{MNPQ}}
\end{equation}
so that the gauge symmetry could after all be that of conventional 6D
diffeomorphisms.\footnote{I would like to thank Paul de Medeiros for discussions about this suggestion.}

Similar considerations apply to the local supersymmetry transformations. In
$D = 5$, the local supersymmetry transformations in a supergravity background
give rise to `background supersymmetry transformations' with symplectic
Majorana spinor parameters $\varepsilon^{\alpha a}$ (where $\alpha$ is a $D =
5$ spinor index and $a = 1, .., 8$ labels the 8 supersymmetries) in which the
gravitino fluctuation $\psi_{\mu}^a$ transforms without a derivative of
$\varepsilon^a$, and `gauge supersymmetries' with spinor parameter
$\varepsilon^{\alpha a}$ under which
\[ \delta \psi^a_{\mu} = \partial_{\mu} \varepsilon^a + \ldots \]
The background symmetries preserving a flat space background form the $D = 5$
super-Poincar{\'e} group. In the free theory, the $D = 5$ super-Poincar{\'e}
symmetry lifts to part of a $D = 6$ super-Poincar{\'e} symmetry with $D = 5$
translation parameters $\xi^{\mu}$ lifting to $D = 6$ ones $\Xi^M$ and
supersymmetry parameters $\varepsilon$ lifting to $D = 6$ spinor parameters
$\hat{\varepsilon}$. The corresponding $D = 6$ supersymmetry charges $Q$ and
momenta $P$ are generators of the (4,0) super-Poincar{\'e} algebra with
\begin{equation} \{Q_{\alpha}^a, Q_{\beta}^b \} = \hspace{0.17em} \Omega^{ab} (\Pi_+
   \Gamma^M C)_{\alpha \beta} P_M  
    \label{fialgss}
    \end{equation}
where $\Pi_{\pm}$ are the chiral projectors
\[ \Pi_{\pm} = \frac{1}{2} (1 \pm \Gamma^7)  \]
$\alpha, \beta$ are $D = 6$ spinor indices and $a, b = 1, \ldots, 8$ are
${USp} (8)$ indices, with $\Omega^{ab}$ the ${USp} (8)$-invariant
anti-symmetric tensor. This is in turn part of the $D = 6$ superconformal
group ${OSp}^{\ast}  (8 / 8)$.

The $D = 5$ gauge symmetries including those with parameters $\xi^{\mu},
\varepsilon^a$ satisfy a local algebra whose global limit is the $D = 5$
Poincar{\'e} algebra, but the $D = 6$ origin of this (at least in the free
theory) is an algebra including the generators $\mathcal{Q}^a_{\alpha M}$ of
the fermionic symmetries with parameter $\varepsilon_N^{\alpha}$ and the
generators $\mathcal{P}^{MNP}$ of the bosonic symmetries with parameter
$\chi^{MNP}$. The global algebra is of the form
\begin{equation}
\{\mathcal{Q}_{\alpha N}^a, \mathcal{Q}_{\beta P}^b \} = \hspace{0.17em}
   \Omega^{ab} (\Pi_+ \Gamma^M C)_{\alpha \beta} \mathcal{P}_{(NP) M}   \label{fialgsss}
   \end{equation}
In the dimensional reduction, the $D = 5$ superalgebra has charges
$Q_{\alpha}^a =\mathcal{Q}_{\alpha 5}^a$, $P_{\mu} =\mathcal{P}_{55 \mu}$.

Supersymmetry provides a further argument against the possibility of a
background metric playing any role in an interacting (4,0) theory in $D = 6$.
The $D = 5$ supergravity can be formulated in an arbitrary supergravity
background, but these cannot be lifted to $D = 6$ (4,0) backgrounds involving
a background metric as there is no (4,0) multiplet including a metric or
graviton. The absence of a (4,0) supergravity multiplet makes problematic the
possibility of a background metric and the standard supersymmetry playing any
role in the $D = 6$ theory. Indeed, the interacting theory (if it exists)
should perhaps be a theory based on something like the algebra   (\ref{fialgsss}) rather than
the super-Poincar{\'e} algebra  (\ref{fialgss}).

\section{Double Copy}

It is intriguing that there is a sense in which gravity can be regarded as a
``square'' of Yang-Mills theory and supergravity can be regarded as a
``square'' of super-Yang-Mills theory. At the level of free theories, there is
a direct construction of supergravity as a square of SYM, with the
supergravity supermultiplet arising from the tensor product of two SYM
multiplets \cite{Borsten:2013bp,
Anastasiou:2013hba,Anastasiou:2014qba}. In the same way, the free (4,0) theory is a square of the free
(2,0) theory \cite{Chiodaroli:2011pp,Borsten:2017jpt}.

Remarkably, for the interacting theories 4D and 5D supergravity scattering
amplitudes can be constructed from ``squaring'' 4D and 5D SYM scattering
amplitudes in the double copy construction \cite{Bern:2010yg}. This suggests another approach to
seeking an interacting (4,0) theory. Given amplitudes for the (2,0) theory,
one could seek to apply the double copy construction to obtain (4,0)
amplitudes. After all, on circle compactification, the (2,0) amplitudes should
reduce to 5D SYM amplitudes and the (4,0) ones should reduce to 5D
supergravity ones, which should in turn be the double copy of the 5D SYM
amplitudes. This approach has been attempted in \cite{Cachazo:2018hqa,Geyer:2018xgb,Albonico:2020mge}.

A major problem with this approach is that we do not know the amplitudes for
the (2,0) theory even at tree level. In  \cite{Cachazo:2018hqa,Geyer:2018xgb,Albonico:2020mge}  formal expressions were derived
that had many of the properties that would be required of (2,0) amplitudes and
which reduced to 5D SYM amplitudes, but they had a number of problems. One was
that they could only be derived for even numbers of external states. Another
was that the poles had problematic non-local residues. As such, the
interpretation of these expressions is unclear. Nonetheless, the double copy
can be applied to these to give formal expressions for the (4,0) theory, but
again it is unclear as to how they might be interpreted and whether they can
be thought of as scattering amplitudes. 

However, as the (2,0) theory exists then it should have well-defined
correlation functions and associated amplitudes, even though we do not yet
understand what these might be. Whatever they are, they should then give (4,0)
amplitudes. As these theories do not have conventional local covariant
interactions, it is to be expected that the corresponding amplitudes should
also involve non-standard features. It will be interesting to see whether
seeking amplitudes could be a useful way to unravel the mysteries of the
interacting theories.

\section{All Fo(u)r Nothing?}

There is a free (4,0) superconformal theory in 6D which provides a conformal
theory of an exotic ``graviton'' but which nevertheless reduces to linearised
5D N=8 supergravity on circle reduction, with a standard graviton. It then
gives a conformal theory of gravity with only two derivatives. Compactifying
on a 2-torus to 4D gives an ${SL} (2, \mathbb{Z})$ symmetry acting
through duality transformations, giving a gravitational duality for the free
theory \cite{Hull:2000zn,Hull:2000rr,Hull:2001iu}. The usual gauge symmetry of gravity is replaced by one with a
parameter which is a 3-tensor instead of a vector field.

The big question is whether there is an interacting version of this theory. It
is of course possible that there is no interacting theory, but it would be
disappointing if M-theory didn't avail itself of the existence of such an
interesting multiplet.

If an interacting theory exists, then it cannot be a conventional theory but
should be some ``non-lagrangian'' theory, presumably of a similar kind to the
(2,0) theory. If there is an interacting (4,0) theory, then the next question
is whether it arises in M-theory. It was argued in \cite{Hull:2000zn} that it should arise as
a strong coupling limit of M-theory toroidally compactified to 5D, understood
as a limit to energies much greater than the Planck scale.

A key question to ask here is whether M-theory compactified to 5D has BPS
states carrying   the singlet central  charge $K$ and, if so, what the spectrum of these is. The
conjecture that the (4,0) theory arises as a strong coupling limit requires
the 5D theory to have BPS states carrying $K$ with the right spectrum to be
Kaluza-Klein modes.
This then provides an important  test for the conjecture  that there could be a 6D (4,0) theory arising as a limit of M-theory:
if M-theory toroidally compactified to 5D does not have  BPS states carrying $K$, then the conjecture is false.
On the other hand,    the presence of    $K$-charged states   with a spectrum consistent with a KK tower  would be strong evidence that 
the conjecture could be true. Importantly, this is a test that can be examined within M-theory and would deepen our understanding of M-theory.

A promising approach to the (4,0) theory is to first understand better amplitudes and correlation
functions for the (2,0) theory and then to use these to construct (4,0)
amplitudes via the double copy mechanism. Such (4,0) tree amplitudes would
then define the classical theory.

There is a further exotic multiplet in 6D with (3,1) supersymmetry \cite{Hull:2000zn}. This
also gives 5D N=8 supergravity on circle reduction and involves a third rank
tensor gauge field with different symmetry properties to that arising in the
(4,0) theory, and the supermultiplet has both vector fields and self-dual
2-form gauge fields. This is not a superconformal supermultiplet and so
requires a dimensionful coupling constant in 6D. Again it is interesting to
ask whether an interacting theory exists and whether it has a role to play in
M-theory. While the (4,0) multiplet arises from the product of two (2,0) ones,
the (3,1) multiplet arises from the product of a (2,0) multiplet with a (1,1)
multiplet. Moreover, (3,1) amplitudes can be obtained from combining (2,0)
with (1,1) amplitudes and in  \cite{Cachazo:2018hqa,Geyer:2018xgb,Albonico:2020mge} formal (3,1) amplitude-like expressions were
found using their (2,0) expressions for ``amplitudes". As in the (4,0) case, these have many
of the properties one might expect from an amplitude but have non-local poles
and other problematic features.

Whether or not the (4,0) theory turns out to have a role to play in M-theory,
it raises some interesting issues. Studying the multiplet led to the duality
between the graviton and dual graviton in linearised gravity \cite{Hull:2000zn,Hull:2000rr,Hull:2001iu}  and to a
gravitational S-duality in 4D \cite{Hull:2000rr}. It raises the possibility that gravity could be
fundamentally conformal and that at high energies it could be described by
some field other than the usual symmetric 2-tensor graviton. More generally,
it is interesting to ask how M-theory behaves at trans-Planckian energies,
especially given the interesting structures found in string theory at
ultra-high energies or string-scale curvatures.

\section*{Acknowledgements}
{This work is supported by the STFC 
Consolidated Grant   ST/T000791/1 and a Royal Society Leverhulme Trust Senior Research
    Fellowship.}


\end{document}